\documentstyle[aps,preprint,epsf]{revtex} 

\begin {document} 

\preprint{\large{\sl Submitted to Phys.\,Rev.\,Lett.}} 

\title{ON THE EXTERNAL FIELD EFFECT IN THE LANDAU THEORY OF  
        THE WEAKLY-FIRST-ORDER PHASE TRANSITION} 
 
\author {M.A.Fradkin\cite{Carl}} 
\address{Institute of Crystallography, Russian Academy  
        of Sciences \\ 59 Leninsky prospect, Moscow 117333 Russia}  
\date{March 13, 1994}  
\draft 

\hyphenation{ma-r-te-n-si-tic} 
\hyphenation{Ae-ro-spa-ce}
\maketitle  

\thispagestyle{empty} 
 
\begin{abstract} 
     The effect of the external field on the weakly-discontinuous  
     first-order phase transition is analyzed in the frame of the Landau 
     theory. The transformation of the free energy expansion as a power series
     in the order parameter is suggested that maps the first-order phase  
     transition onto the second-order one under the 'effective' external  
     field, that depends on both temperature and on real field value. The  
     presence of the third-degree term in the Ginzburg-Landau expansion is  
     shown to preserve the weakly-discontinuous phase transition for some  
     values of the external field in contrast with the second-order phase  
     transition. The case of proper ferroelastic (martensitic) phase  
     transition from cubic lattice to tetragonal one is considered and the  
     dependence of the transition temperature on the external hydrostatic  
     as well as uniaxial pressure is found. 
\end{abstract}
 
\bigskip

        It is well-known that the continuous (second-order) phase transition 
        disappears under the applied external field conjugated to the  
        order-parameter. In other word, there is no phase transition when  
        the temperature is varying under the fixed non-zero external  
        field\cite{LandLif,Toledano,Levanyuk}. For the systems exhibiting  
        weakly-discontinuous first-order phase transition in the absence of  
        the external field, different behavior can occur when the field is  
        applied, because the transition takes place between co-existing phases 	 
	and sufficiently small field can only change the energy of phases but  
        can not make them unstable. Thus, the field conjugated to the order  
        parameter may shift the temperature of the first-order phase  
        transition instead of suppressing the transition itself.  
 
        In the present Letter transformation of the Ginzburg-Landau
        expansion of free energy as a power series in the single-component 
        order parameter is suggested that maps the first-order phase transition 
        onto the second-order one under the 'effective' external field. It is
        shown that real first-order phase transition is described by the 
	transformed functional, when this 'effective' field vanishes. In 
	such a case the transition without the symmetry breaking takes place 
	between two different minima of the free energy, both are corresponding 
	to the low-symmetry phases with different values of the order 
	parameter, so the transitions remains to be of the fist order. The 
	ferroelastic phase transition\cite{Salje} from cubic lattice to 
	tetragonal one provides an example of this situation with respect to 
	spontaneous strain as an order parameter and uniaxial pressure as 
	conjugated external field.
 
        The Landau theory of continuous phase transitions\cite{LandLif}  
        suggests that general expression for the (Ginzburg-Landau) expansion  
        of the difference in Gibbs free energy between the phases has the  
        following form 
\begin{equation} 
        \Delta {\cal G} = {\alpha \over 2}(T - T_c)\eta^2 + {C \over 4}\eta^4 
        \label{GL0} 
\end{equation} 
        where $T_c$ is a critical temperature and the coefficients $\alpha$  
        and $C$ should be positive. The equilibrium value of the order  
        parameter $\eta$ is determined by the minimization of $\Delta  
        {\cal G}$ with respect to $\eta$: 
\begin{equation} 
        \frac{\partial \Delta {\cal G}}{\partial \eta} = 0  
        \hspace{12pt} {\rm and} \hspace{12pt} \ \ \ 
        \frac{\partial^2 \Delta {\cal G}}{\partial^2 \eta} > 0. 
        \label{minimum} 
\end{equation} 
        The solutions are the high-symmetry phase with $\eta = 0$, stable for 
        $T > T_c$ and low-temperature phase with $\eta^2 = 
        {\frac{\alpha}{C}}(T_c-T)$ that is stable for $T < T_c$. The  
        equilibrium phases do not co-exist and the dependence of $\eta$ on $T$         	
	is continuous in the critical point, hence this model describes the  
        second-order phase transition.  
 
        Upon introducing the effect of external field $E$, conjugated to the  
        order-parameter $\eta$, the free energy can be written in the  
        following way 
\begin{equation} 
        \Delta \tilde{\cal G} = {\tau \over 2} \ \eta^2 + {\eta^4 \over 4} 
                - \sigma \eta \ , 
        \label{secondfield}  
\end{equation} 
        via convenient units for the field $\sigma = E/C$, temperature  
        $\tau = \alpha (T - T_c) / C$ and energy $\Delta \tilde{\cal G} =  
        \Delta {\cal G}/C$. 
        The condition (\ref{minimum}) leads to the cubic equation 
\begin{equation} 
         \eta^3 + \tau \eta - \sigma = 0  
        \label{min1f} 
\end{equation} 
        with discriminant  
\begin{equation} 
         Q = \bigl({\tau \over 3}\bigr)^3 + \bigl({\sigma \over 2}\bigr)^2 .  
\end{equation} 
        For any value of external field $\sigma$ the 
        high-symmetry phase with $\eta = 0$ no longer provides the stable  
        solution of Eq.(\ref{min1f}). Instead, one get $\eta \neq 0$ 
        for any temperature.  
 
\begin{figure} 
\epsffile{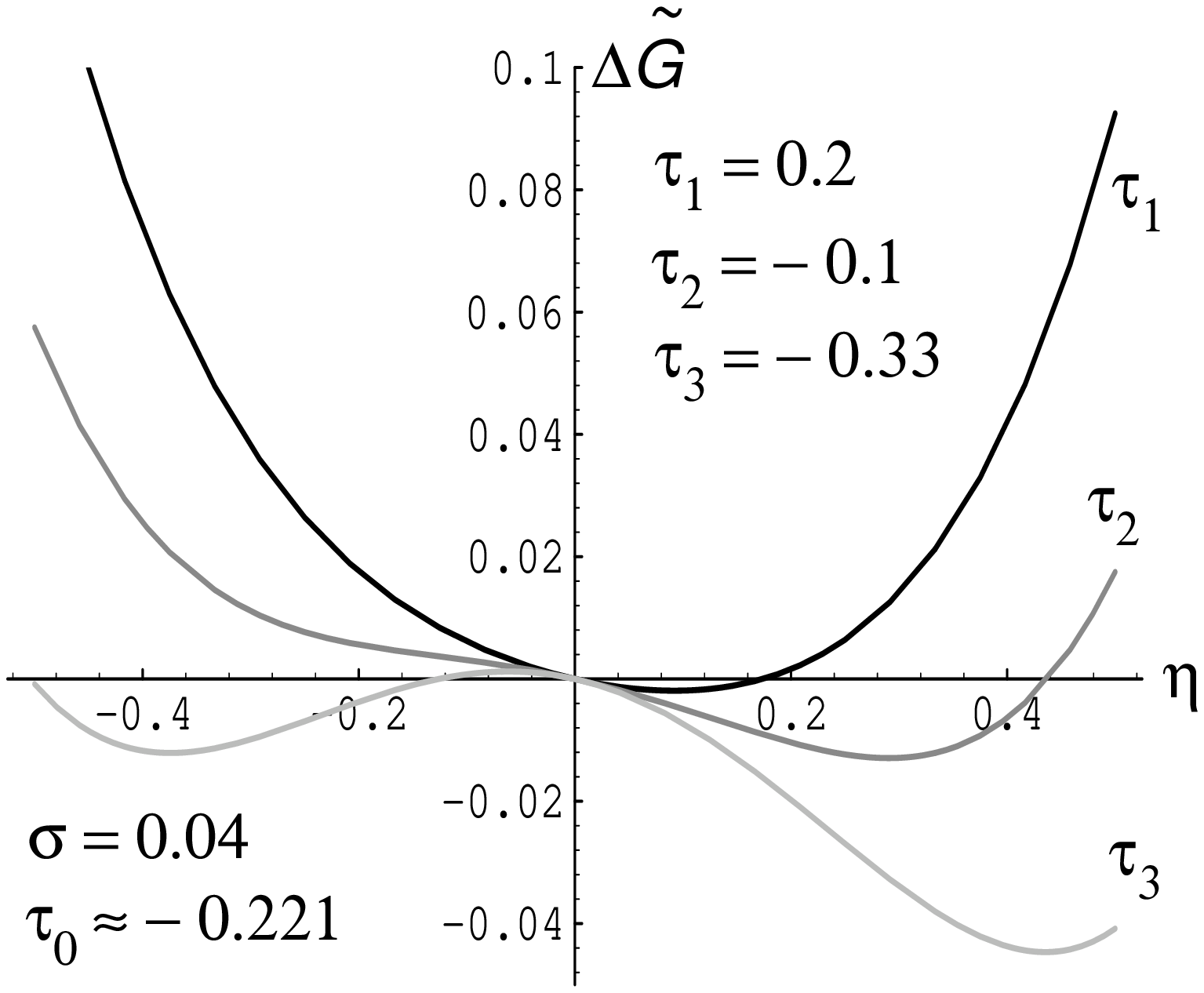} 
\caption{The dependence of the Gibbs energy on the order parameter $\eta$  
        under the applied field for different  
        temperatures $\tau_1 > \tau_2 > \tau_0 > \tau_3$ in the case of  
        the second-order phase transition.}  
\label{IIenergy} 
\end{figure}

\begin{figure} 
\epsffile{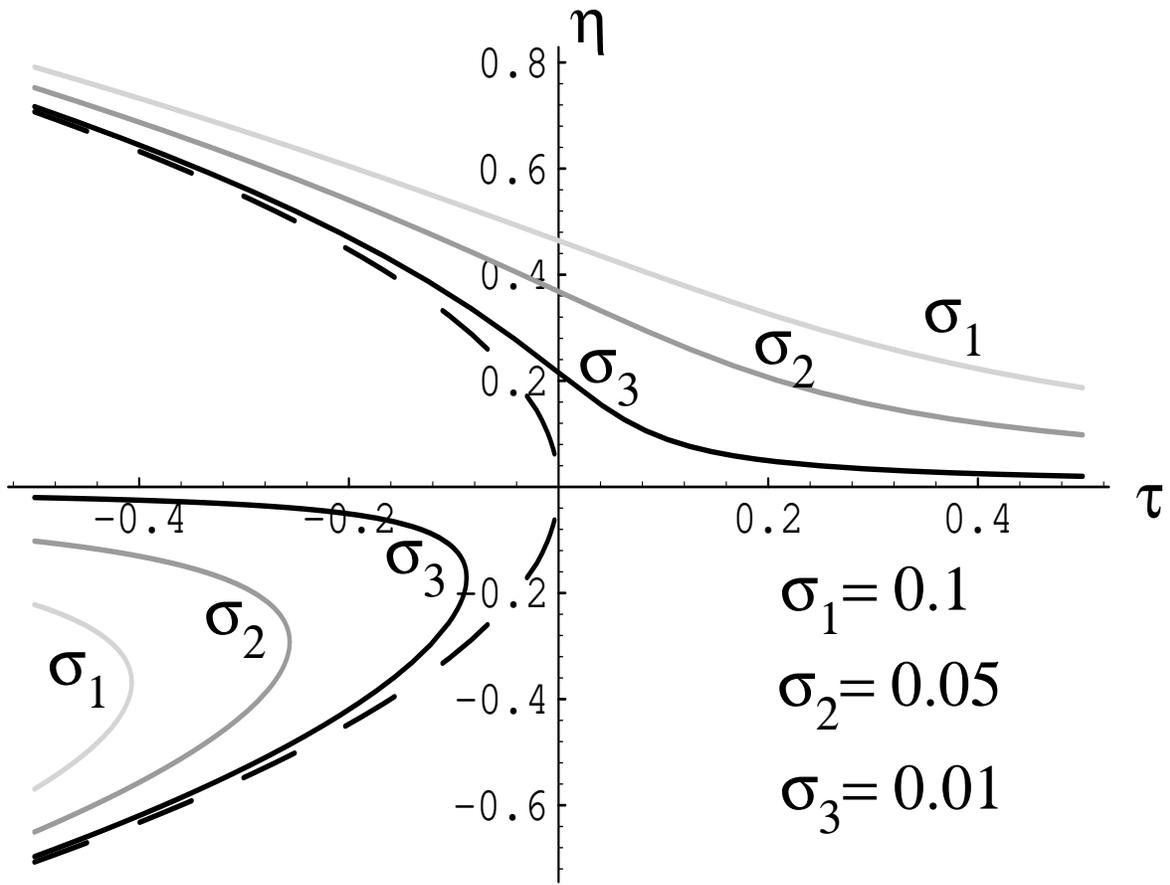} 
\caption{The order parameter dependence on the temperature in various fields 
        for the case of the second-order phase transition. Dashed line  
        corresponds to the absence of external field, $\sigma = 0$.}   
\label{IIordpar} 
\end{figure}

        The cubic equation is known\cite{Korn} to have one solution in real  
        numbers for $Q > 0$ and three ones for $Q < 0$. Thus, the additional  
        minimum of the Gibbs free energy appears for  
\begin{equation} 
        \tau \le \tau_0 = - 3 \ \bigl({\sigma \over 2}\bigr)^{2 \over 3}. 
\end{equation} 
        The $\Delta \tilde{\cal G}  
        (\eta)$ curves and the dependencies of $\eta$ on temperature in  
        different fields are shown in the Fig.\ref{IIenergy} and   
        Fig.\ref{IIordpar}. 
 
        It is seen from the Fig.\ref{IIenergy} and might be proven rigorously 
        that different minima of the $\Delta \tilde{\cal G} (\eta)$ curve have  
        different energies for any temperature $\tau < \tau_0$. As the second  
        minimum appears at $\tau_0$ with the free energy of ${3 \over 4}  
        (\sigma /2)^{4 \over 3}$ which is higher than that of existing  
        high-temperature phase, $-6 (\sigma /2)^{4 \over 3}$, the latter  
        state remains stable throughout all the region of  
        the phase co-existence. Only the condition of $\sigma = 0$ through  
        the degeneracy with respect to sign of $\eta$ implies the equal 
        energies of different minima and leads to a phase transition of 
        the first order when the external field goes through zero at fixed  
        temperature $\tau < 0$. In other words, the variation of temperature  
        and external field act in a different way on the systems  
        describing by the Ginzburg-Landau expansion (\ref{GL0}),  
        because the field variation may lead to the phase change but the  
        temperature one may not. 
 
        The weakly-first-order phase transition arises in the Landau  
        theory when the symmetry of the system allows to have third-degree  
        invariant composed by the order-parameter component. Hence,  
        corresponding term should be included in the Ginzburg-Landau  
        expansion: 
\begin{equation} 
        \Delta {\cal G} = {\alpha \over 2}(T - T_c)\eta^2 + 
                {B \over 3}\eta^3  + {C \over 4}\eta^4. 
        \label{TOT} 
\end{equation} 
        The physical interest has a case of $B < 0$ and the free energy  
        expansion can be written in the form 
\begin{equation} 
        \Delta \tilde{\cal G} = {\frac{C^3}{B^4}} \Delta {\cal G} = 
                {\tau \over 2} \ \zeta^2 - {\zeta^3 \over 3} 
                + {\zeta^4 \over 4} - \sigma \zeta \ ,  
        \label{firstfield} 
\end{equation} 
        with $\tau =\alpha C (T - T_c)/B^2, \ \eta = - (B/C) \zeta \ $ and  
        $\sigma = - (C^2/B^3) E$. 
 
        For the $\sigma = 0$ case, Eq.(\ref{minimum}) gives two possible  
        minima of the free energy, those are undistorted phase with  
        $\zeta = 0$ unstable for $\tau < 0$ and low-symmetry one with  
        $\zeta = (1 + \sqrt{1 - 4 \tau})/2$ appearing for $\tau < 1/4$.  
        Hence, some region of the phase co-existence appears, and  
        the phase energies become equal at $\tau_\ast = 2/9$, though  
        the supercooling of the high-temperature state as well as superheating  
        of the low-temperature one are possible. The jump in order parameter  
        is $\Delta \zeta = {2/3}$ and the nucleation  
        potential barrier has a height ${\cal E}_b = \Delta \tilde{\cal G}  
        (\zeta_b,\tau_\ast) = {1 \over 324}$. It means that the first-order 
	phase transition takes place at the temperature $\tau_\ast$. 
 
        Substituting $\zeta = \tilde{\zeta} + {1/3}$ into the  
        Ginzburg-Landau expansion (\ref{firstfield}), we get the  
        third-order term excluded 
\begin{equation} 
        \Delta \tilde{\cal G} =  
                {\tilde{\tau} \over 2} \ \tilde{\zeta}^2 +  
                {\tilde{\zeta^4} \over 4} - \tilde{\sigma} \tilde{\zeta} + 
                \Delta \tilde{\cal G}_0\ ,  
        \label{transgl} 
\end{equation} 
        where 
\begin{displaymath} 
        \tilde{\tau} = \tau - {1 \over 3} \ ; \ \  
	\tilde{\sigma} = \sigma - {\tau \over 3} + {2 \over 27} \ \ {\rm and} 
	\ \ \Delta \tilde{\cal G}_0\ = {{\tau} \over 18} - 
	{{\sigma} \over 3} - {1 \over 108}. 
\end{displaymath} 

        This is equivalent to the free energy expansion (\ref{secondfield})  
        for the second-order phase transition under the external field, the  
        only difference consisting of the term $\Delta \tilde{\cal G}_0$ that  
        is independent on $\tilde{\zeta}$. It appears  
        because the free energy is counted with respect to the $(\zeta = 0)$  
        or $(\tilde{\zeta} = - {1/3})$ state, that implies  
        $\Delta \tilde{\cal G}_0\ = \Delta \tilde{\cal G}(\tilde{\zeta} = 0) 
        \ne 0$.  
         
        The condition (\ref{minimum}) leads to the cubic equation  
(\ref{min1f})    
        with the temperature $\tau$ and field $\sigma$ replaced by the  
        effective ones $\tilde{\tau}$ and $\tilde{\sigma}$, respectively.  
        The sign of discriminant  
\begin{equation} 
        Q = \bigl({\tilde{\tau} \over 3}\bigr)^3 +  
                \bigl({\tilde{\sigma} \over 2}\bigr)^2  
                \propto  4 \sigma + 27 \sigma^2 -  
                        18 \sigma \tau - {\tau^2} + 4 \tau^3  
\end{equation} 
        of this equation indicates, whether it has one root or three ones in  
        the real numbers. The latter case corresponds to the appearance of  
        different minima on $\Delta \tilde{\cal G} (\tilde{\zeta})$,  
        second minima of the free energy appearing when $Q(\tau, \sigma) < 0$.  
 
        Hence, (\ref{transgl}) can be considered as the Ginzburg-Landau  
        expansion for the phase transition between the states, 
        related with different minima of the Gibbs free energy. The  
        minima have non-zero values of order parameter, because the symmetry is  
        broken already by the applied field for any temperature. As there is  
        no symmetry breaking, it is not a true phase transition, described by  
        the Landau theory, however, the undistorted phase with $\zeta = 0$ can  
        be treated as an analog of ideal high-symmetry   
        "praphase"\cite{Levanyuk} that would allow the symmetry  
        reduction to both of the phases which provide minima of the free  
        energy. 
 
\begin{figure} 
\epsffile{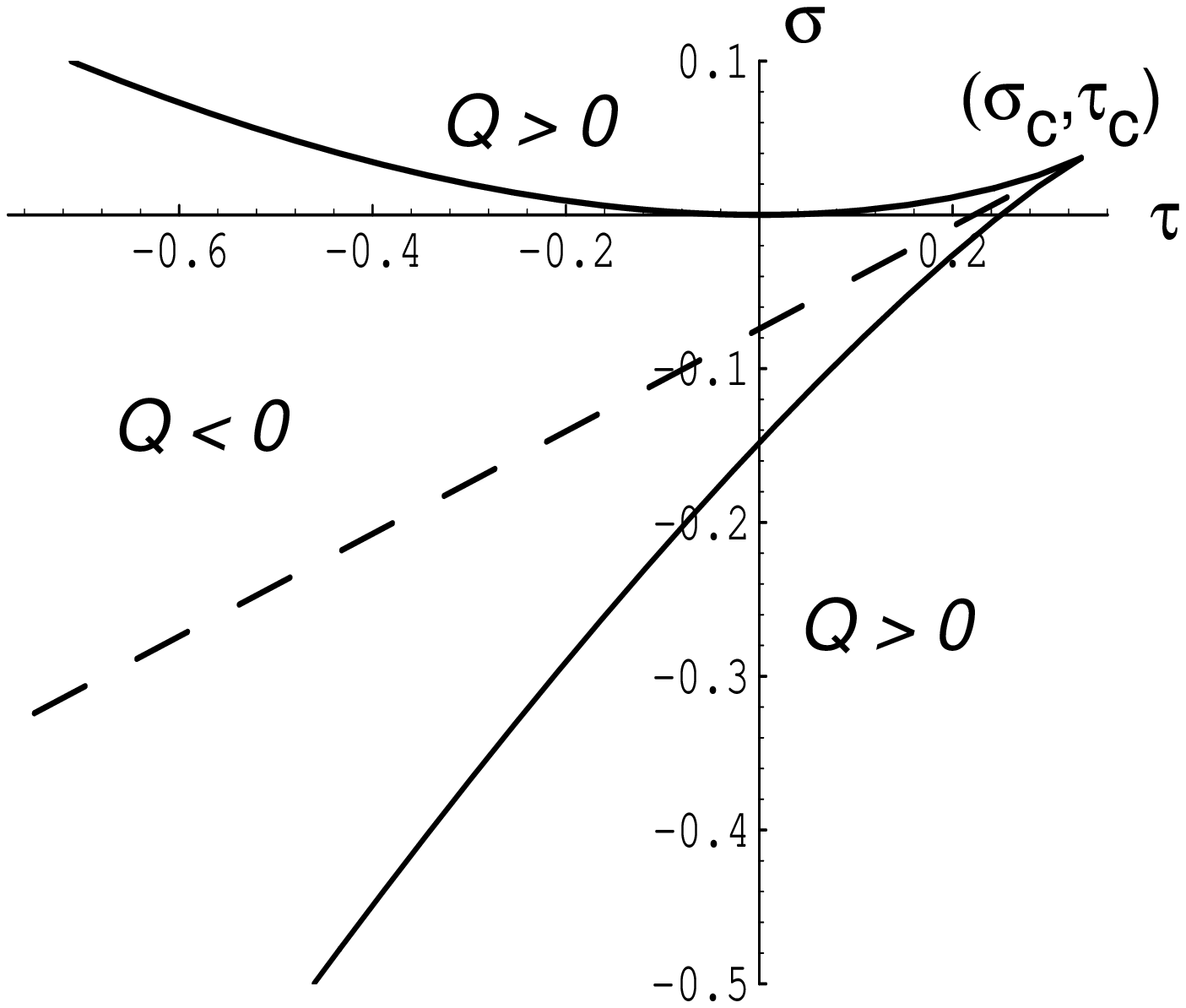} 
\caption{The region of the phase coexistence. The dashed line corresponds to   
        points of the first-order phase transition. It terminates in the  
        critical point $(\tau_c = {1/3}, \sigma_c = {1/27})$.}  
        \label{diagram}  
\end{figure}

        The phase diagram in $(\tau, \sigma)$ plane is shown at the  
        Fig.\ref{diagram}. Additional minimum of the free energy appears for  
        $\sigma_1 \le \sigma \le \sigma_2$ with 
\begin{equation} 
        \sigma_{1,2} = - {2 \over 27} \left( 1 \pm ( 1 - 3\, \tau )^{3\over 2}  
        \right) + {\tau \over 3},  
\end{equation}   
        that leads to the hysteresis with respect to the external  
        field $\Delta \sigma = (4 \left( 1 - 3\, \tau \right)^{3\over 2})/27$. 
         
        According to an analogy with the second-order phase  
        transition described by (\ref{secondfield}),  
        the different minima of the $\Delta \tilde{\cal G}(\tilde{\zeta})$  
        have equal energy only at $\tilde{\sigma} = 0$. This is the condition 
        of the first-order phase transition and it determines the effect of  
        applied field on transition temperature $\tau_\ast$  
\begin{equation} 
        \tau_\ast(\sigma) = 3\, \sigma + {2/9} \ . 
        \label{tastsigma} 
\end{equation} 
        For $\sigma = 0$ we get naturally $\tau_\ast(0) = {2/ 9}$. The  
        Eq.(\ref{tastsigma}) corresponds to the straight line on $(\tau,  
        \sigma)$ plane (Fig.\ref{diagram}). For $\tau > {1/ 
        3}$ on this line the equilibrium phase has $\tilde{\zeta} = 0$ or  
        $\zeta = {1/3}$. This state is an analog of the undistorted 
        high-symmetry phase of the Landau theory without an external field  
        which is unstable for $\tau < {1/3}$.  
        Below $\tau = {1/3}$ on the $\tau_\ast(\sigma)$ line it becomes  
        a maximum of $\Delta \tilde{\cal G}$, located between  
        two minima with $\tilde{\zeta}_{1,2} = \pm \sqrt{- \tilde{\tau}}$,  
        separated by the energy barrier and the order parameter jump  
\begin{displaymath} 
        {\cal E}_b = {9 \over 4} \sigma^2 - {\sigma \over 6} + {1 \over 324}  
        \ \ \ {\rm and} \ \ \ \Delta \zeta = {2 \over 3} \sqrt{1 - 27 \sigma}, 
\end{displaymath} 
        respectively 
 
        As the line of first-order transition in the $(\tau, \sigma)$  
        phase diagram separates states without symmetry-breaking  
        relation\cite{LandLif}, it terminates in  
        critical point $(\tau_c = {1/3}, \sigma_c = {1/27})$, which  
        is an analog of the continuous phase transition from the state with  
        $\tilde{\zeta} = 0$. The jump in order parameter as well as the  
        potential barrier vanish at the critical point, hence, the  
        weakly-first-order phase transition disappears for $\sigma > \sigma_c$  
        or $\tau > \tau_c$. In a contrast with the second-order case where  
        arbitrary small external field destroys the phase transition, here  
        we find that the transition is preserved in the fields lower than  
        $\sigma_c$.  
 
	Let us consider the proper ferroelastic phase transition as an example. 
	The free energy difference between the phases is due to spontaneous 
	strain and can be described by the Ginzburg-Landau expansion of the 
	elastic energy as a power series in the strain 	
	components\cite{Salje,Bocc}. The second-degree term in this expansion 
	is a linear combination of the second-order elastic constants that 
	vanishes at the critical temperature\cite{Cowley}. It is an eigenvalue 
	of the lattice stiffness matrix corresponding to the relevant 
	irreducible representation of the symmetry group of high-temperature 
	phase $G$\cite{LandLif} and the strain tensor components transforming 
	with respect to this representation form the order 
	parameter\cite{Bocc}. If the symmetry of low-temperature 
	phase in known {\em a-priory}, than a particular combination of the 
	strain tensor components that is invariant under an action of the 
	symmetry group $G_1$ of the low-symmetry phase may be treated as a 
	single-component order parameter. Due to long-range nature of the 
	elastic forces in solids the Landau theory appears to be 
	applicable up to the transition temperature\cite{Levanyuk,Cowley}.
 
        In the case of ferroelastic phase transition from cubic to tetragonal 
        lattice the symmetry allows the third-order term to occur in the  
        Ginzburg-Landau expansion, but it appears to be small in In-Tl,  
        V$_3$Si and some other alloys\cite{Brass}. Hence, this phase  
        transition should be weakly discontinuous and belongs to the  
        "soft-mode" class, because it is accompanied by a softening of the  
        corresponding acoustic mode of atomic vibrations\cite{Cowley}. The  
        Ginzburg-Landau expansion of the elastic free energy up to the fourth  
        order\cite{Liak} is written in terms of a suitable symmetrized  
        combination of the strain components that describes relevant symmetry  
        breaking\cite{Bocc} 
\begin{equation} 
        \eta_1 = ( - \epsilon_{xx} - \epsilon_{yy} 
                + 2 \epsilon_{zz})/{\sqrt{6}}  
        \label{ordpar} 
\end{equation} 
        and corresponds to the extension along $z$ axis without the volume  
        change. The elastic free energy can be  
        written in the form of Eq.\ref{TOT} with the following  
        elastic constant combinations as coefficients\cite{Liak} 
\begin{eqnarray} 
        {\alpha_1}(T - T_c) = (C_{11} - C_{12}) \nonumber \\ 
        B_1 = ( C_{111} - 3 C_{112} + 2 C_{123})/{2 \sqrt{6}} \nonumber \\ 
        C_1 = ( C_{1111} - 4 C_{1112} + 3 C_{1122})/12 \nonumber  
\end{eqnarray} 
 
        There might be an additional degree of freedom that arises  
        from the volume change\cite{Good} 
\begin{equation} 
	\eta_0 = ( \epsilon_{xx} + \epsilon_{yy} + \epsilon_{zz})/\sqrt{3}
\end{equation} 
 	and contributes the coupling  
        term $\Delta {\cal G}_{int} = D \eta_0 \eta_1^2$ with $D =  
        (C_{111} - C_{123})/(2 \sqrt{3})$. This coupling implies the finite  
        volume change at the first-order transition. The minimization of  
        $\Delta {\cal G}(\eta_0, \eta_1)$ with respect to $\eta_0$ leads to  
        renormed coefficients 
\begin{displaymath} 
        \alpha = \alpha_1 + 2 \alpha_0 D \ \ \ {\rm and} \ \ \ \ \ 
         C = C_1  - {{2 D^2}/A_0} 
\end{displaymath} 
        of the second and fourth orders, with $\alpha_0$ and $A_0$ being the  
        volume thermal expansion coefficient and bulk modulus,  
        respectively.  
 
        An applied pressure gives rise the 'external' stress tensor $\hat{E}$  
        corresponding to linear term $ - \hat{\epsilon} \hat{E}$ in the  
        elastic energy\cite{LanLifEl}. The tetragonal symmetry of low  
        temperature phase, determined by the order parameter (\ref{ordpar})  
        can only be conserved under $\hat{E}$ having a form 
\begin{equation} 
        E_{i,k} = \delta_{i,k} E_{i,i} \ ; \ \ \ \ \ \ \ E_{xx} = E_{yy}, 
\end{equation} 
        of superposition of applied uniaxial pressure along $z$ axis 
\begin{displaymath} 
        E = ( - E_{xx} - E_{yy} + 2 E_{zz})/{\sqrt{6}}  
\end{displaymath} 
        with hydrostatic one $P = {\rm Tr} (\hat{E})/{\sqrt{3}}$. Thus,  
        E is the external field, conjugated to the primary order parameter  
        $\eta_1$, whereas hydrostatic pressure is shown\cite{Fradkin} to  
        shift the critical temperature of ferroelastic phase transition and,  
	hence, leads to the change of $\tau$ at fixed temperature 
\begin{equation} 
        \tau \rightarrow \tau - {{2 D C} \over {A_0 B^2}} \ P. 
\end{equation} 
         
        The line of the first-order phase transition on the $(T,P,E)$ phase  
        diagram has a form 
\begin{equation} 
        T_{\ast} = T_c + {{2 B^2} \over {9 \alpha C}} +  
        {2 D \over {\alpha A_0}} \,P - {3 C \over {\alpha B}} \,E 
\end{equation} 
        with the critical values of temperature $T_c^{\prime} = T_c + B^2/(3 
	\alpha C)$ and uniaxial pressure $E_c = - B^3/(27 C^2)$. 
        The critical hydrostatic pressure depends linearly on the temperature  
\begin{equation} 
        P_c (T) = {{A_0} \over {2 D}} \left(\alpha (T - T_c) -  
                {{B^2} \over {3 C}}\right) 
\end{equation} 
        and vanishes when $T$ goes to $T_c^{\prime}$. 
 
        To conclude, I have considered the transformation of the  
        Ginzburg-Landau expansion of the free energy for the case of the  
        weakly-discontinuous first-order phase transition and have shown  
        that it is equivalent to the free energy expansion for the  
        second-order phase transition with included effect of the external 
	conjugated field. It is shown that in a contrast with the second-order  
        case where arbitrary small external field destroys the phase  
        transition, the first-order transition is preserved in the fields  
        lower than $\sigma_c$. The effect of the external uniaxial as well as  
        hydrostatic pressure upon the proper ferroelastic (martensitic) phase  
        transition from cubic to tetragonal phase is considered and the  
        equation on the values of the transition temperature, hydrostatic  
        pressure and uniaxial one is derived.  
 
\acknowledgments 
 
        This work was supported, in part, by a Soros Foundation Grant awarded  
        by the American Physical Society as well as by International Science  
        Foundation Grant. The final part of the work has been done at Carleton  
        University under the hospitality of Prof.J.Goldak.

\epsfxsize=6.5in  

\begin{thebibliography}{99} 
 
  \bibitem[\dagger]{Carl} 
        Present address: Department of Mechanical 
        and Aerospace Engineering, Carleton University, Ottawa, Ont., 
        K1S 5B6, Canada; {\em e-mail:} {\tt mfradkin@next.mrco.carleton.ca}
 
\bibitem{LandLif} L.D. Landau, E.M. Lifshitz, {\em Statistical Physics}, 3rd  
        edition, Oxford, Pergamon, 1981. 
 
\bibitem{Toledano} J.C. Toledano, P.Toledano, {\em  
        The Landau theory of phase  
        transitions}, World Scientific, Singapore, 1987. 
 
\bibitem{Levanyuk} A.P. Levaniuk, A.S. Sigov, {\em Defects and structural  
        phase transitions}, Gordon and Breach, New York, 1988. 
 
\bibitem{Korn}  G.A. Korn, {\em Mathematical handbook for scientists 
           and engineers},  2d ed.,  1968 
 
\bibitem{Salje} E.K.H. Salje, {\em Phase transitions in ferroelastic  
        and co-elastic crystals}, Cambridge University Press, Cambridge, 1990. 
 
\bibitem{Bocc} N. Boccara, 
        Ann. Phys. (N.Y.) {\bf 40}, p.40, (1968). 
 
\bibitem{Cowley} R.A. Cowley, Phys. Rev. B {\bf 13}, p.4877 (1976). 
 
\bibitem{Brass} M.P.Brassington, G.A. Sounders,  
        Proc. Roy. Soc. {\bf A387}, p.289, (1983). 
 
\bibitem{Liak} J. Liakos, G.A. Sounders, 
        Phil. Mag. A {\bf 46}, p.217, (1982). 
 
\bibitem{Good} R.J. Gooding, Y.Y. Ye, C.T. Chan, K.M. Ho, B.N. Harmon, 
        Phys. Rev. B {\bf 43}, p.13626 (1991).   
 
\bibitem{Fradkin} M.A. Fradkin, Preprint MAF -- 12/93, (1993). 
 
\bibitem{LanLifEl} L.D. Landau, E.M. Lifshitz, 
        {\em Theory of Elasticity}, 3rd edition, Oxford, Pergamon, 1981. 
 
\end{thebibliography}
\end{document}